# Rapid Lung Ultrasound COVID-19 Severity Scoring with Resource-Efficient Deep Feature Extraction

Pierre Raillard[1,2], Lorenzo Cristoni[3], Andrew Walden[4], Roberto Lazzari[5], Thomas Pulimood[6,7], Louis Grandjean[8,9], Claudia AM Gandini Wheeler-Kingshott[10,11], Yipeng Hu[1,2], Zachary MC Baum[1,2]

[1] Centre for Medical Image Computing, University College London
[2] Wellcome/EPSRC Centre for Surgical & Interventional Sciences, University College London
[3] Frimley Park Hospital, Frimley Health NHS Foundation Trust
[4] Royal Berkshire Hospital, Royal Berkshire NHS Foundation Trust
[5] Hospital de La Santa Creu I Sant Pau, Barcelona
[6] West Suffolk Hospital, West Suffolk NHS Foundation Trust
[7] Cambridge University Hospital, University of Cambridge
[8] Great Ormond Street Children's Hospital NHS Foundation Trust
[9] Institute of Child Health, University College London
[10] NMR Research Unit, Queen Square MS Centre, UCL Queen Square Institute of Neurology
[11] Department of Brain and Behavioural Sciences, University of Pavia
zachary.baum.19@ucl.ac.uk

**Abstract.** Artificial intelligence-based analysis of lung ultrasound imaging has been demonstrated as an effective technique for rapid diagnostic decision support throughout the COVID-19 pandemic. However, such techniques can require days- or weeks-long training processes and hyper-parameter tuning to develop intelligent deep learning image analysis models. This work focuses on leveraging 'off-the-shelf' pre-trained models as deep feature extractors for scoring disease severity with minimal training time. We propose using pre-trained initializations of existing methods ahead of simple and compact neural networks to reduce reliance on computational capacity. This reduction of computational capacity is of critical importance in time-limited or resource-constrained circumstances, such as the early stages of a pandemic. On a dataset of 49 patients, comprising over 20,000 images, we demonstrate that the use of existing methods as feature extractors results in the effective classification of COVID-19-related pneumonia severity while requiring only minutes of training time. Our methods can achieve an accuracy of over 0.93 on a 4-level severity score scale and provides comparable per-patient region and global scores compared to expert annotated ground truths. These results demonstrate the capability for rapid deployment and use of such minimally-adapted methods for progress monitoring, patient stratification and management in clinical practice for COVID-19 patients, and potentially in other respiratory diseases.

# 1 Introduction

The use of point of care ultrasound (POCUS) has grown increasingly along with the evidence of its use in improving patient outcomes. POCUS has been considered a convenient, effective and useful imaging technique for triaging and monitoring and diagnosis of COVID-19 patients [1–3]. These approaches do not require the movement of patients to different locations given ultrasound equipment may be used at the bedside, and have been increasingly used for lung ultrasound imaging (LUS) throughout the COVID-19 pandemic [4]. However, despite the increased use of LUS, the unprecedented number of patients with COVID-19 at times of the pandemic caused an oversaturation of the diagnostic capacities in some hospitals [5]. This is likely due to the inherent challenges with ultrasound, whereby experienced operators may be required to obtain images of sufficient quality, and requisite knowledge is then required to properly interpret the acquired images. Given the pressures on healthcare systems, several studies have investigated the use of deep learning to assist clinicians in triage, diagnosis, severity scoring, and monitoring COVID-19 patients [6–13]. These methods have been useful in the stratification of COVID-19 patients by aiding in determining the location of pathological image features or biomarkers, all based on LUS data. Of note, these works typically involve the development of complete end-to-end training processes for custom, or tailored versions of state-of-the-art architectures.

The requirement for end-to-end development and training processes is ever-present in existing works which seek to address the problem of scoring COVID-19 severity from LUS. Methods in the literature focus on the using complete end-to-end training processes [7, 14] or the use of transfer learning and fine-tuning [15]. In practice, this results in a classification method that relies on large volumes of retrospectively labelled data, as is the case for the work we present here. However, as these aforementioned methods rely on a complete end-to-end training or extensive fine-tuning protocol, they also require a large amount of compute time and potentially require thorough experimentation to optimize performance for use in practice. This presents real-world challenges, especially and most importantly; obtaining high-quality labelled data [16]. This challenge may be most pressing at the beginning of an epidemic or pandemic, where imaging from infected patients may be scarce for training performant models. Furthermore, obtaining expert labels, or taking days or weeks to train and tune numerous hyper-parameters ahead of validating a model for clinical use could be increasingly detrimental during the peak of an outbreak. In these instances, the use of existing pre-trained models could reduce the time needed to deploy diagnostic supports.

By leveraging 'off-the-shelf' state-of-the-art network architectures, we may explore utilizing existing initializations, learned for large-scale visual recognition tasks, such as the ImageNet Challenge [17], for deep feature extraction prior to classification. Through the use of these networks for feature extraction, we may reduce the training process from the order of weeks or days to minutes, reducing the reliance on computational capacity; something of critical importance in time-limited, or resource-and-expertise constrained circumstances. As such, in this work, we study the effectiveness of using various existing pre-trained network architectures for the classification of COVID-19 severity scores.

## 2 Methods

We consider multiple state-of-the-art convolutional neural network architectures for classifying the severity of COVID-19-related pneumonia in LUS images of healthy and COVID-19-positive patients, with diagnosis confirmed by PCR tests. We utilize the Xception [18], ResNet50 [19], and VGG16 [20] networks as 'off-the-shelf', pre-trained classification models, which we can fine-tune using our labelled data, comprising of four severity classes, as described in Section 2.3.

### 2.1 Severity Classification and Tested Networks

Several commonly used neural networks for image classification, Xception [18], ResNet50 [19], and VGG16 [20], were adapted to automatically identify severity scores in the LUS images of COVID-19 positive and healthy patients. At inference, these networks will then predict whether a given image belongs to one of four severity classes, indicating which visible features are likely to be present, as well as an estimate of the severity of the disease in a patient. In using well-established network architectures, we may direct our investigation towards the feasibility of ultrasound image classification on a small-scale dataset using pre-trained feature extractors.

### 2.2 Implementation Details

Each of the neural networks were implemented in Tensorflow [21] and Keras [22]. Reference quality, open-source code and pre-trained ImageNet weights from Keras were adopted for the Xception, ResNet50, and VGG16 networks. ImageNet weights were utilized to prevent the need for pre-training, reducing the computational requirements for this work and preventing the need to rely on existing ultrasound datasets. While existing ultrasound datasets may have been used to pre-train the feature extractors, this would require the existence of large, robust, diverse, and well labelled ultrasound datasets, as well as the computational capacity to train such networks to be able to appropriately utilize the weights. The fully-connected layers at the top of each network were replaced with a Dropout layer; with a probability of 0.5, and a fully-connected layer with 4 neurons; one per severity class added. The final layer used a SoftMax activation to determine the probability of membership in each class for a given image. All pre-trained layers were frozen before training. Data augmentation with rotation, shifting, scaling and horizontal flipping was used during training. All models were trained for 3 epochs on an Intel i7 CPU, with a mini-batch size of 32. A cross-entropy loss was employed with a learning rate of 0.001 using the Adam optimizer [23].

### 2.3 Data

In this work, US images were captured at two different hospitals by two clinicians using a Butterfly iQ US probe (Butterfly Inc., Guilford, CT, USA). Images were captured from 37 COVID-19 positive patients admitted to hospital and undergoing treatment with COVID-19-related pneumonitis and 12 healthy patients, yielding 23732 images. All 23732 images were manually assigned an image-level severity score label by a biomedical engineering student familiar with LUS. Labels were reviewed and verified by experienced US imaging researchers with five or more years of experience with clinical US imaging. The severity score assigned to each image is adopted from the Lung Ultrasound Score [24] protocol for rating the severity of the disease in a given LUS image. The Lung Ultrasound Score has been shown to provide strong correlation with lung density, as measured by quantitative computed tomography scans [24]. The Lung Ultrasound Score protocol places an image into one of four different categories, Score 0, Score 1, Score 2, and Score 3 – as defined in Table 1. Using the Lung Ultrasound Score protocol to label our images; we obtain 10865 Score 0 images, 2013 Score 1 images, 10514 Score 2 images, and 340 Score 3 images. As the COVID-19 positive patient images used in this work were acquired from subjects which were hospitalized at time of acquisition, we suspect the distribution of Scores in our data may not directly correspond with the distribution of Scores seen among all COVID-19 cases, giving rise to a large prevalence of Score 2 images in our dataset. Example images for each severity score label are provided in Figure 1.

**Table 1.** Definition of severity level and key associated features of the Lung Ultrasound Score.

| Severity Score | Visible Features |
| --- | --- |
| Score 0 | A-lines with at most two B-lines<br>With or without irregularities to the pleural line |
| Score 1 | Irregularities to the pleural line<br>Visibility of vertical artifacts on at most 50% of the pleura |
| Score 2 | Broken pleural line, pleura damaged or irregular<br>Artefacts on at most 50% of the pleura<br>Consolidation areas visible |
| Score 3 | Broadly visible and dense white lung<br>With or without large consolidation visible<br>Tissue-like patterns |

Each category corresponds to a progressive loss of aeration, wherein Score 0 indicates normal aeration, and Score 3 indicates a complete loss of aeration. To give a complete score for a patient, 12 regions are observed per patient, with 6 per hemithorax as superior and inferior anterior, lateral, and posterior. The global per patient score is the sum of the individual region scores, giving a range of 0 to 36. The regional and global scores have yielded strong correlations with measurements of lung density from computerized tomography imaging [25–26].

### 2.4 Experiments

To evaluate the effectiveness of each model for severity score classification, we perform a 3-fold cross-validation where only the added fully-connected layer is trainable during fine-tuning, leveraging the learned weights of the pre-existing layers to detect relevant features in the images. The COVID-19 and healthy datasets were split into the three cross-validation sets, on a patient level. This splitting ensured that no images from the same patient were found within different folds of the training and testing datasets. Additionally, we strived to maintain approximately similar sizes of the datasets between each fold. Each fold contains 12 or 13 COVID-19 positive patients and 4 healthy patients.

For each network, we perform all folds of the cross-validation, wherein we train for 3 epochs, with most layers frozen; as previously described in Section 2.2. To evaluate each network, we evaluate performance based on accuracy, as well as precision, recall, and the area under the receiver operating characteristic curve (ROC-AUC) score for each class. We additionally present training time and a retrospective analysis of the predicted Lung Ultrasound Score at a patient level based on aggregate frame-level classifications.

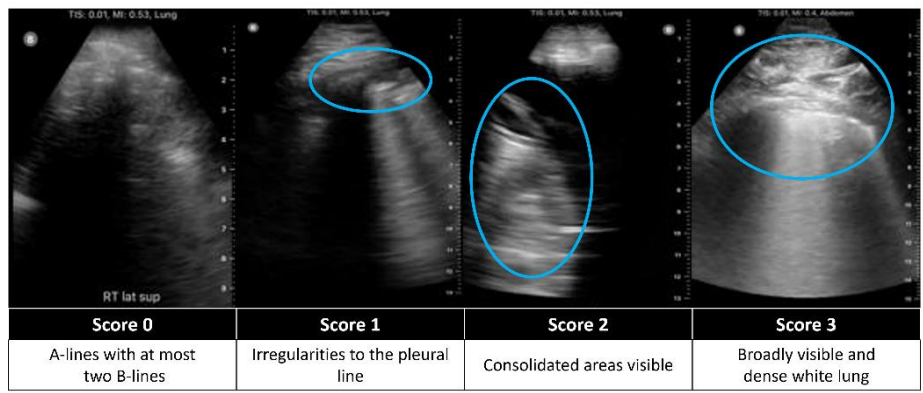

**Fig. 1.** Multiple sample LUS images and their corresponding severity classification. Blue circles indicate identifiable regions of severity in each presented image.

## 3     Results

Overall, an accuracy of 0.948, 0.935, and 0.942 were obtained for the Xception, ResNet50, and VGG16 networks, respectively, on a frame-based classification. This is comparable to the classification performance seen in end-to-end trained networks for severity classification, where accuracy has been demonstrated at 0.976 using similar architectures and a similar 4-tiered scoring system [15]. Region-level accuracies for each of the networks are presented in Figure 2. Table 2 summarizes the macro-precision, macro-recall, ROC-AUC scores and mean training time on CPU for each network. While our methods are trained approximately 30-60 minutes on CPU, existing end-to-end trained methods may take approximately 90 minutes in multi-GPU training schemes with a similarly-sized dataset [15]. The confusion matrices given in Figure 3 summarizes the per-class predictive performance of each of the networks, averaged over all 3 folds of the cross-validation. Figure 3 also presents the ROC curves per severity score class for each method. While the ResNet50 architecture exhibits the highest performance and lowest mean training time of the three methods tested, there is no significant difference between the accuracy, precision, and recall reported for all pairwise comparisons between the three methods.

Lastly, we present a retrospective analysis wherein we determine the complete Lung Ultrasound Score per patient, to yield a global diagnostic score. In each instance, the predicted patient was a part of the test set for the given round of cross-validation. We indicate the number of ground truth images in each region per score, relative to the number predicted, whereby the ground truth and predicted region scores corresponds to the Lung Ultrasound Score with the highest number of images (Figure 4).

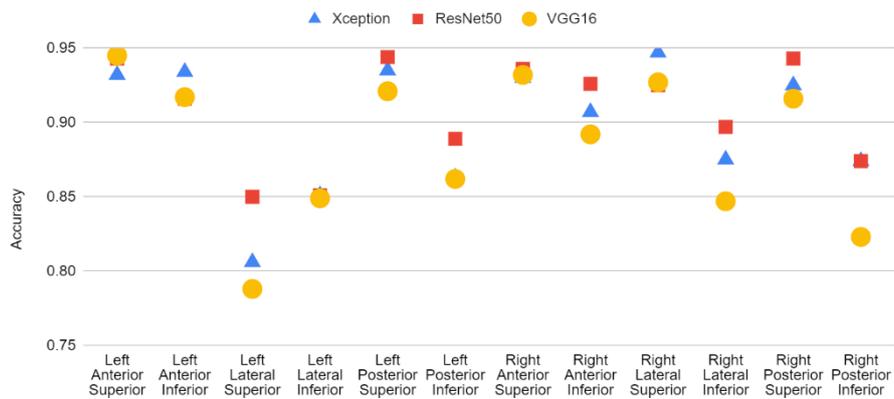

**Fig. 2.** Summary of region-level accuracies, giving the predictive scores over the 3 fold cross-validation for the Xception, ResNet50, and VGG16 networks.

**Table 2.** Summary of cross-validation evaluation metrics for each network architecture.

| Network Architecture | Macro-Precision | Macro-Recall | ROC-AUC score | Mean Training Time (min) |
|---|---|---|---|---|
| Xception | 0.839 | 0.812 | 0.978 | 33.92 |
| ResNet50 | 0.859 | 0.819 | 0.991 | 25.97 |
| VGG16 | 0.811 | 0.737 | 0.971 | 73.67 |

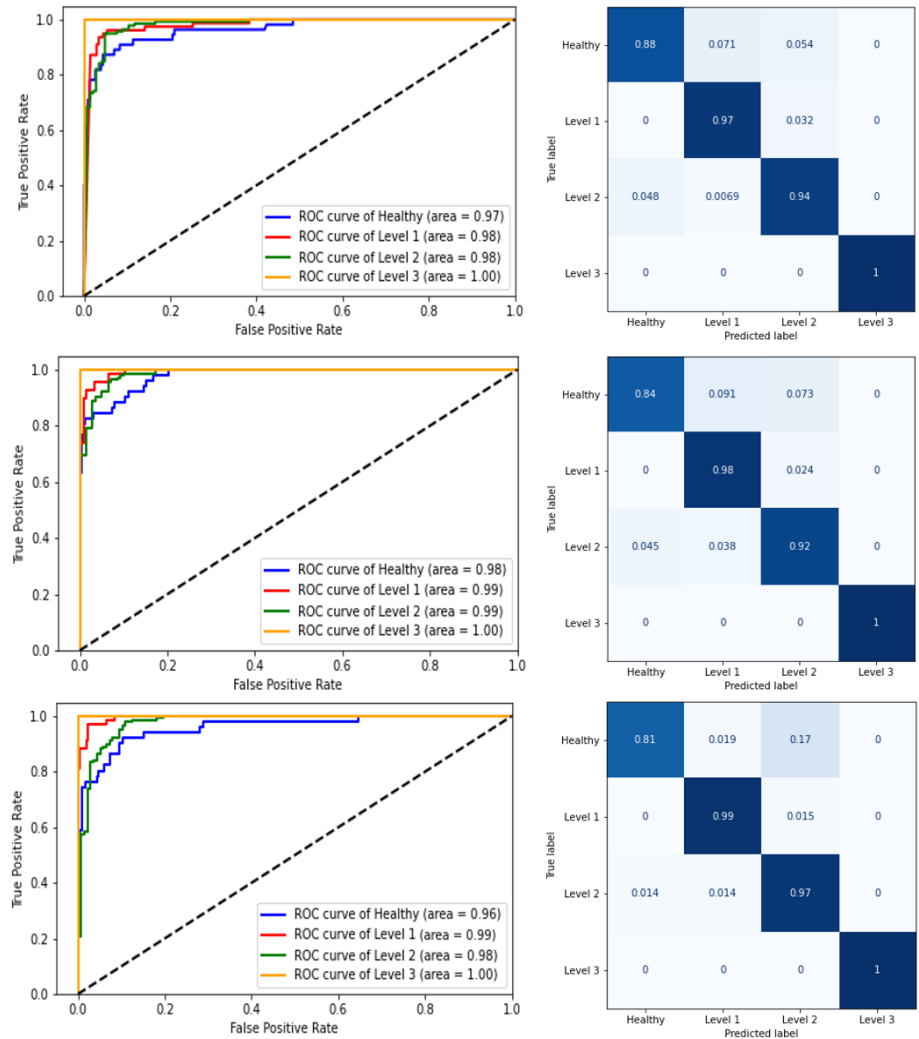

**Fig. 3.** Receiver operating characteristic curves (left) and confusion matrices (right) summarizing predictive scores over the 3 fold cross-validation for the Xception (top), ResNet50 (middle), and VGG16 (bottom) networks. Confusion matrices values are normalized across the ground-truth class label.

## 4    Discussion and Conclusion

This work presents the validation of classifying COVID-19 pneumonitis severity scores through deep feature extraction via state-of-the-art deep learning architectures on real clinical LUS data, acquired from COVID-19 patients. While most work in severity scoring explores applying a complete training process from randomly initialized weights, or a useful initialization, we demonstrate the effectiveness of training only the final layers of the network, and relying on the pre-training process to derive effective feature extraction; something of critical importance in time-limited, or resource-constrained circumstances. In particular, we report that by training only the final layer of a network, initialized with weights learned from the ImageNet dataset, we may obtain acceptable accuracy, even with a limited set of patient data. This demonstrates the effectiveness of these pre-learned feature extractors when applied to new data, even in a different domain. These results provide evidence in support of the usefulness of applying simpler, and more efficient training methods to existing architectures for applications with limited data in medical imaging applications. Further studies are required to determine if such methods are practical and effective for confounding lung conditions, such as interstitial lung disease, pneumonia, or other pathologies which cause sub-pleural changes as well.

| Region | Score 0 | | Score 1 | | Score 2 | | Score 3 | | Region Score | | Global Score | |
|---|---|---|---|---|---|---|---|---|---|---|---|---|
| | Ground Truth | Predicted | Ground Truth | Predicted | Ground Truth | Predicted | Ground Truth | Predicted | Ground Truth | Predicted | Ground Truth | Predicted |
| Left Anterior Superior | 5 | 2 | 44 | 45 | | 2 | | | 1 | 1 | 9 | 9 |
| Left Anterior Inferior | 6 | 12 | 60 | 55 | | 1 | | | 1 | 1 | | |
| Left Lateral Superior | 31 | 24 | 30 | 35 | | | | | 0 | 1 | | |
| Left Lateral Inferior | 35 | 38 | 39 | 36 | | | | | 1 | 0 | | |
| Left Posterior Superior | 65 | 67 | 2 | | | | | | 0 | 0 | | |
| Left Posterior Inferior | | 3 | 65 | 62 | | | | | 1 | 1 | | |
| Right Anterior Superior | 21 | 20 | 36 | 35 | | 2 | | | 1 | 1 | | |
| Right Anterior Inferior | 2 | 1 | 26 | 29 | 14 | 12 | | | 1 | 1 | | |
| Right Lateral Superior | 13 | 15 | 42 | 37 | 3 | 6 | | | 1 | 1 | | |
| Right Lateral Inferior | | 1 | 54 | 51 | 2 | 4 | | | 1 | 1 | | |
| Right Posterior Superior | 47 | 55 | 23 | 15 | | | | | 0 | 0 | | |
| Right Posterior Inferior | 11 | 4 | 44 | 54 | 3 | | | | 1 | 1 | | |

| Region | Score 0 | | Score 1 | | Score 2 | | Score 3 | | Region Score | | Global Score | |
|---|---|---|---|---|---|---|---|---|---|---|---|---|
| | Ground Truth | Predicted | Ground Truth | Predicted | Ground Truth | Predicted | Ground Truth | Predicted | Ground Truth | Predicted | Ground Truth | Predicted |
| Left Anterior Superior | 3 | 6 | 24 | 26 | 32 | 27 | | | 2 | 2 | 16 | 15 |
| Left Anterior Inferior | | 2 | 29 | 27 | | | | | 1 | 1 | | |
| Left Lateral Superior | | | 42 | 47 | 7 | 2 | | | 1 | 1 | | |
| Left Lateral Inferior | | 4 | 39 | 35 | | | | | 1 | 1 | | |
| Left Posterior Superior | | | 5 | 9 | 27 | 23 | 2 | 2 | 2 | 2 | | |
| Left Posterior Inferior | 5 | 8 | 54 | 49 | | 2 | | | 1 | 1 | | |
| Right Anterior Superior | | | 42 | 46 | 6 | 2 | | | 1 | 1 | | |
| Right Anterior Inferior | | | 27 | 35 | 37 | 29 | 6 | 6 | 2 | 1 | | |
| Right Lateral Superior | | | 34 | 37 | 13 | 10 | | | 1 | 1 | | |
| Right Lateral Inferior | | 4 | 54 | 61 | 22 | 11 | | | 1 | 1 | | |
| Right Posterior Superior | 7 | 16 | 104 | 95 | 3 | 3 | | | 1 | 1 | | |
| Right Posterior Inferior | | | 39 | 37 | 57 | 59 | 5 | 5 | 2 | 2 | | |

**Fig. 4.** Retrospective analysis of per-zone severity classifications on two sample patients from our datasets. Values in the ground truth column indicate the number of images which appear in that severity score per zone, per patient. Values in the predicted column indicate the number of images which appear in that severity score per zone, per patient. Zones with the highest value are selected as the severity score for a zone in the ground truth and predicted label. Zones which the ground truth and predicted severity correspond are indicated in green, zones which differ are indicated in red.

**Acknowledgments.** This work is supported by the Wellcome/EPSRC Centre for Interventional and Surgical Sciences (203145Z/16/Z). C.A.M. Gandini Wheeler-Kingshott is supported by the MS Society (#77), Wings for Life (#169111), Horizon2020 (CDS-QUAMRI, #634541), BRC (#BRC704/CAP/CGW), and allocation from the UCL QR Global Challenges Research Fund (GCRF). Z.M.C. Baum is supported by the Natural Sciences and Engineering Research Council of Canada Postgraduate Scholarships-Doctoral Program, and the University College London Overseas and Graduate Research Scholarships.